\documentclass[manuscript]{aastex}

\makeatletter
\let\@dates\relax
\makeatother

\usepackage{graphicx}
\usepackage{rotating}
\usepackage{pdflscape}
\usepackage{color}
\usepackage{lineno}
\linenumbers*[1]
\usepackage{mathtools}
\usepackage{amsmath}
\usepackage{float}
\usepackage{color}

\shorttitle{Flux tubes in the solar wind}
\shortauthors{Arnold et al.}

\begin{document}
\title{Observation of flux tube crossings in the solar wind}

\author{L. Arnold$^1$, G. Li$^{1,*}$, X. Li$^1$, Y. Yan$^2$}
\affil{
$^1$ Department of Physics and CSPAR, University of Alabama in Huntsville 35899, USA, gang.li@uah.edu \\
$^2$ Key Laboratory of Solar Activity, National Astronomical Observatories of Chinese Academy of Sciences,
Beijing 100012, China \\  
 \quad \\
Published: The Astrophysical Journal, 766, 2,  2013, \\
doi:10.1088/0004-637X/766/1/2
}

\begin{abstract}
Current sheets are ubiquitous in the solar wind. 
They are a major source of the solar wind MHD turbulence intermittency. 
They may result from non-linear interactions of the solar wind MHD turbulence
or are the boundaries of flux tubes that originate from the solar surface. 
Some current sheets appear in pairs and are the boundaries of 
transient structures such as magnetic holes and reconnection exhausts, or
the edges of pulsed Alfv\'{e}n waves. For an individual
current sheet, discerning whether it is a flux tube boundary or due to non-linear interactions, or 
the boundary of a transient structure is difficult. In this work, using data from 
the {\sl Wind} spacecraft, we identify two three-current-sheet events. Detailed examination of these two events 
suggest that they are best explained by the flux tube crossing scenario. Our study provides a convincing 
evidence supporting the scenario that the solar wind consists of flux tubes where
distinct plasmas reside. 
\end{abstract}

\maketitle

\section{Introduction}
\par
	 
The solar wind provides a natural environment to study MHD 
turbulence in a collisionless plasma. Over the past few decades the launches of spacecraft such 
as Voyager, Helios, Ulysses, {\sl Wind}, and ACE have made available a significant amount of data for 
analyzing the solar wind MHD turbulence. 

The first theory of hydrodynamic turbulence, suggested by \citet{Kolmogorov41}, known as the 
K41 theory, predicted a magnetic field power-law spectrum $\sim k^{-5/3}$. 
This $-5/3$ exponent arises from the nonlinear 
interactions of the homogeneous hydrodynamic turbulence in which energy is cascaded from large scales to 
small scales. For incompressible MHD turbulence where 
the cascading process is mediated by counter-propagating Alfv\'en wave packets, 
the Iroshnikov-Kraichnan (IK) theory \citep{Iroshnikov64, Kraichnan65} and some recent theories of strong 
MHD turbulence \citep{Boldyrev06, Boldyrev.Perez09} predict a power spectrum  $\sim k^{-3/2}$.

One important concept in turbulence is intermittency. \citet{Ruzmaikin.etal95} 
suggested that intermittent structures can affect the solar wind MHD turbulence power spectrum.
They pointed out that the effect of intermittency in the 
solar wind MHD turbulence is to reduce the exponent of the power law spectrum.
\citet{Ruzmaikin.etal95} further suggested that intermittency is
``in the form of ropes, sheets or more complicated fractal forms.'' 
Recently, in studying current sheets (CS) in the solar wind, \citet{Li.etal11} found that the power 
spectrum of solar wind magnetic field behaves as K41 in periods that have abundant numbers of 
current sheets and behaves as IK in periods that are almost current-sheet free (see also \citet{Borovsky10}). 
Since these current sheets are a source of intermittency, the study of \citet{Li.etal11} supports 
\citet{Ruzmaikin.etal95}. 
	
A current sheet is a 2D structure across which the magnetic field direction changes abruptly.
Current sheets can be of large scales. For example, the heliospheric current sheet and current sheets 
found in CME-driven shocks are all large scale current sheets. These are not the subjects of this study.
Here we consider current sheets that are of small scales.

Some current sheets occur in pairs. 
These can be tangential discontinuities (TDs), often forming the two boundaries of a magnetic hole 
(see the review of \citet{Tsurutani.etal11} and references therein), or 
rotational discontinuities (RDs) which are the boundaries of an exhaust from a reconnection site 
(see \citep{Gosling.etal05} and the review of \citet{Gosling11}). 
Comparing to magnetic holes, reconnection exhausts can be of larger scales. 
\citet{Gosling07} has found that the typical width of a reconnection exhaust is $\sim 10^4$ km and
some reconnection exhausts can be as wide as $10^5$ km. 
Consequently, these boundaries may be practically identified as ``single-current-sheet'' event.
 
Most current sheets do not occur in pairs.  These current sheets can be generated through non-linear 
interactions in the MHD turbulence \citep{Zhou.etal04, Chang.etal04}. 
Using ACE data, \citet{Vasquez.etal07} examined magnetic field discontinuities which can have very
 small spread angles for Bartels rotation 2286 (day 7 to 33 in 2001). They found that the statistical 
properties of these discontinuities form a single population and they are consistent with turbulence 
generated in-situ. By examining the probability density functions (PDF) 
of the magnetic field components from a 1D spectral code,  \citet{Greco.etal08} showed that 
current sheets often occur at the super Gaussian tail of the PDF. Moreover,  
\citet{Greco.etal09} found that, at the inertial scale, in which the energy cascading rate is 
independent of the scale, the PDF of waiting times (WTs) between MHD discontinuities that are identified 
in the solar wind using the method of \citet{Tsurutani.Smith79} and those from MHD simulations are 
very similar, suggesting that these structures can be explained as a natural result of the non-linear 
interaction of the solar wind MHD turbulence.

Other opinions exist.
In an earlier work, \citet{Bruno.etal01} studied current sheets in the solar wind by analyzing 
Helios 2 data using
the minimum variance method to show how the magnetic field changed over selected time periods. 
\citet{Bruno.etal01} were the first to suggest that these structures may be boundaries between flux tubes. 
\citet{Borovsky08} analyzed an extended time period of magnetic field from the ACE spacecraft and 
examined the distribution of the spread angle across the current sheets. He showed that 
the angle distribution has two populations and suggested that the second population, 
dominating at large angles, could be ``magnetic walls'' and originate from the surface of the Sun. 
A solar wind that consists of many flux tubes can be viewed as a structured solar wind.  
In both the work of \citet{Bruno.etal01} and \citet{Borovsky08}, the solar wind is envisioned to 
be full of structures, the flux tubes.  Observed at a spacecraft, these structures are convected 
out with the solar wind. A similar scenario where structures convecting out from the Sun has been 
proposed by \citet{Tu.Marsch91}. Analysis on the cross helicity $\sigma_c$ and residual 
energy $\sigma_r$ by \cite{Bruno.etal07} supported the proposal of \citet{Tu.Marsch91}.

Regardless the origin of a current sheet, \citet{Li08} developed a procedure to systematically 
identify these structures.
Using this procedure, \citet{Li.etal08} examined current sheets in the solar wind and in 
the Earth's magnetotail using {\it Cluster} magnetic field data and concluded that current sheets are more 
abundant in the solar wind.  Later, \citet{Miao.etal11} examined over 3 years' worth slow wind data 
using {\it Ulysses} observations and found there were $2$ populations for the distribution of 
the spread angle across current sheets, in agreement with \citet{Borovsky08}. 

Perhaps a large fraction of current sheets identified in the solar wind are due to the non-linear 
interactions of the solar wind MHD turbulence, 
as shown in the work of \citet{Greco.etal08, Greco.etal09}.
However,  a statistical study such as  \citet{Greco.etal08, Greco.etal09} 
can not rule out the possibility that  
some current sheets in the solar wind are boundaries of flux tubes. 
{  Indeed, \citet{Borovsky08} has used plasma data including proton density and temperature, Helium 
abundance, electron strahl strength, etc. to identify possible plasma boundaries. Plasma data, however, 
is often of lower time resolution than magnetic field data.
Furthermore, plasmas in different flux tubes may have similar properties except different velocities
and magnetic field directions. Therefore, to unambiguously separate these two populations can be hard. }
Note, the occurrence rates of these two populations may have different radial dependence and/or 
different solar wind type dependence. 

\begin{figure}[ht]
\includegraphics[width=0.9\textwidth]{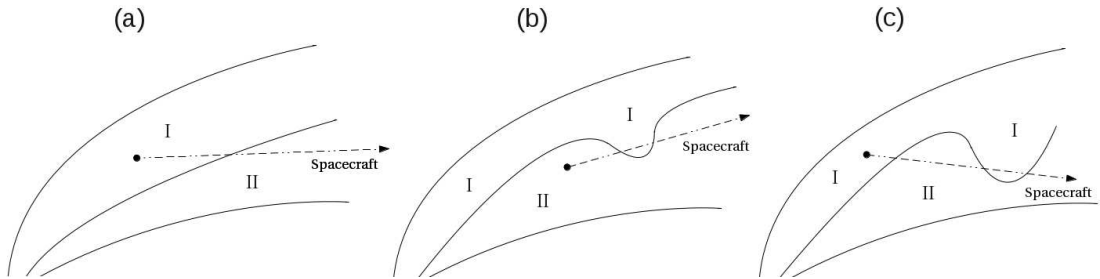}
\caption{Cartoons illustrating (a)  a single-current-sheet event,  (b) a 
double-current-sheet event, and (c) a triple-current-sheet event. Note that in the case of  
triple-current-sheet event, the spacecraft traverses through two distinct plasmas in the sequence of 
``I, II, I, II''. Consequently, we expect to find the plasma properties to vary accordingly.
Dashed line with arrow represents the relative trajectory of the {\it Wind} spacecraft passing through 
the flux tubes.}
\label{Cartoon}
\end{figure}

In this work, as an effort to identify flux tubes in the solar wind,  we present a case study of 
two ``triple-current-sheet'' event using data from spacecraft {\it Wind}.
 A triple-current-sheet event is where three current sheets occurred in a relatively short period of time. 
The reason that we want to search for a triple-current-sheet event is the following.
In the flux-tube scenario, the solar wind plasmas reside in different flux tubes and the solar wind magnetic 
field and plasma properties differ in these flux tubes. 
Since flux tubes are 3D structures, we expect the boundary between two adjacent flux tubes to be 
curved and have small-scale ripples. This is shown in the cartoons in Figure~\ref{Cartoon}. 
As these flux tubes are convected out past a spacecraft, 
depending on the relative configuration of the spacecraft trajectory and these ripples,  
one expects to observe most often a single crossing as in Figure~\ref{Cartoon} (a), and sometimes 
a double crossing as in Figure~\ref{Cartoon} (b), and occasionally a triple crossing as in 
Figure~\ref{Cartoon} (c).
These three different cases are referred to as
 ``single-current-sheet'' events, ``double-current-sheet'' events, and ``triple-current-sheet'' events in
 this study.

A triple-current-sheet event  can be used to discriminate between the scenario where current sheets are 
generated in-situ and the scenario where current sheets originate from the surface of the Sun.
In the former case, one expects no correlations between these current sheets in the sense that
 plasmas before and after these current sheet crossings need not show any relationships. 
In the latter case, however,  the spacecraft traverses through two distinct plasmas 
in the sequence of ``I, II, I, II'', so the observed plasma properties do not vary arbitrarily.

\section{Data Selection and Analysis}
We use the $3$ s plasma and magnetic field data from the 
 3DP (\citet{Lin.etal95}) and magnetic field (\citet{Lepping.etal95}) experiments
on the {\it Wind} spacecraft.
The data period was from September to October of 1995, which was during the declining phase of the 
solar cycle. It is ideal to select data in the solar minimum period due to lack of transient 
structures such as  CMEs.  
For the data analysis method of current sheet identification, the readers are referred to \citet{Li08} and 
\citet{Miao.etal11}.  

In the following, we first present a single-current-sheet event and a double-current-sheet event. We
then present two triple-current-sheet events.
		
\begin{figure}[ht]
\includegraphics[width=0.9\textwidth]{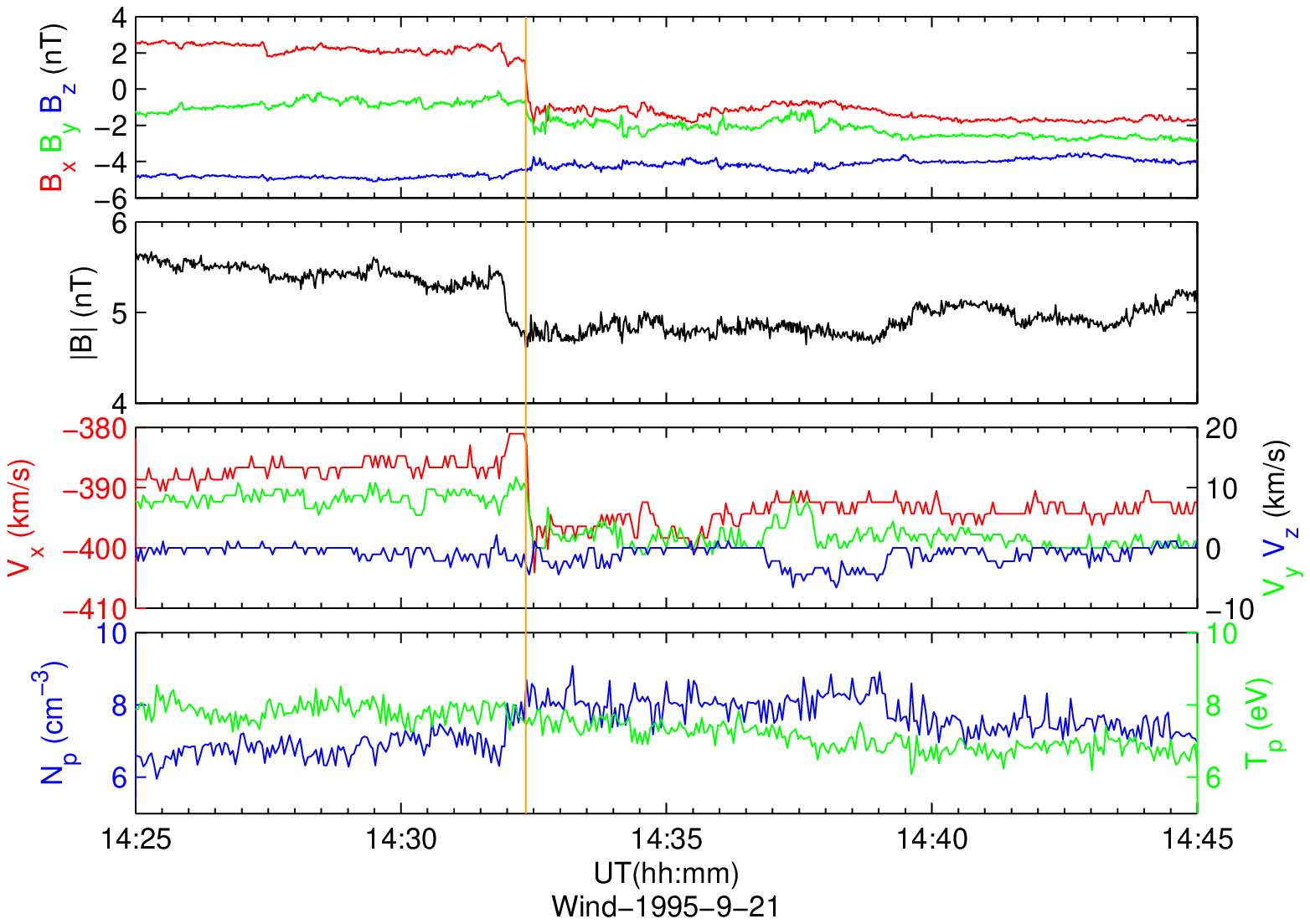}
\caption{A single current sheet event occurred in 1995-09-21. Shown from top to bottom are 
the $3$ components of the vector magnetic field in the Geocentric Solar Ecliptic (GSE)
coordinate system, the magnitude of magnetic field, the  $3$ components of the vector proton velocity 
in the Geocentric Solar Ecliptic (GSE) coordinate system, and the solar wind proton number density and proton 
temperature, respectively. The brown vertical line marks the location of the current sheet.}

\label{AnglePlot1}
\end{figure}

Figure~\ref{AnglePlot1} is a single-current-sheet event which occurred on 1995-09-21.
The current sheet in Figure~\ref{AnglePlot1} is located at 14:32 UT and is shown by the brown vertical line.
 Before the current sheet the magnetic field magnitude $|B|$ decreases and the proton density $N_p$ increases.
In the scenario where a current sheet is the boundary  of a flux tube,
these changes across the current sheet occur because plasmas in different flux tubes have different 
properties (as shown in panel~(a) of Figure~\ref{Cartoon}). 
However, one need not invoke the flux-tube-crossing scenario to explain Figure~\ref{AnglePlot1}. 
It can be simply a tangential discontinuity (TD) or one side of a reconnection exhaust.
Indeed, careful examination shows that there was another small current sheet on $\sim 14:32$ UT, when 
$|B|$ decreased and $N_p$ increased. 
Our selection procedure did not pick out this earlier current sheet.

The change across the current sheets are Alfv\'{e}nic. 
The angle between $\delta \vec{B}$ and $\delta \vec{V}$ across the 
current sheet is $7^{\circ}$.  For the earlier current sheet (that did not get picked up by our procedure), the
 angle is $173^{\circ}$.  Such parallel and anti-parallel Alfv\'{e}nic changes are always associated with 
a reconnection exhaust \citep{Gosling11}. 
Furthermore, there was also a decrease of magnetic field and an increase of number density (but not temperature) 
between these two current sheets, providing another support for identification of a reconnection exhaust.
 {Therefore,  Figure~\ref{AnglePlot1}, although identified as a single-current-sheet event using the 
\citet{Li08} algorithm, is really part of a pair of a bifurcated current 
sheet \citep{Gosling.etal05, Gosling07, Gosling11}.}

\begin{figure}[ht]
\includegraphics[width=0.9\textwidth]{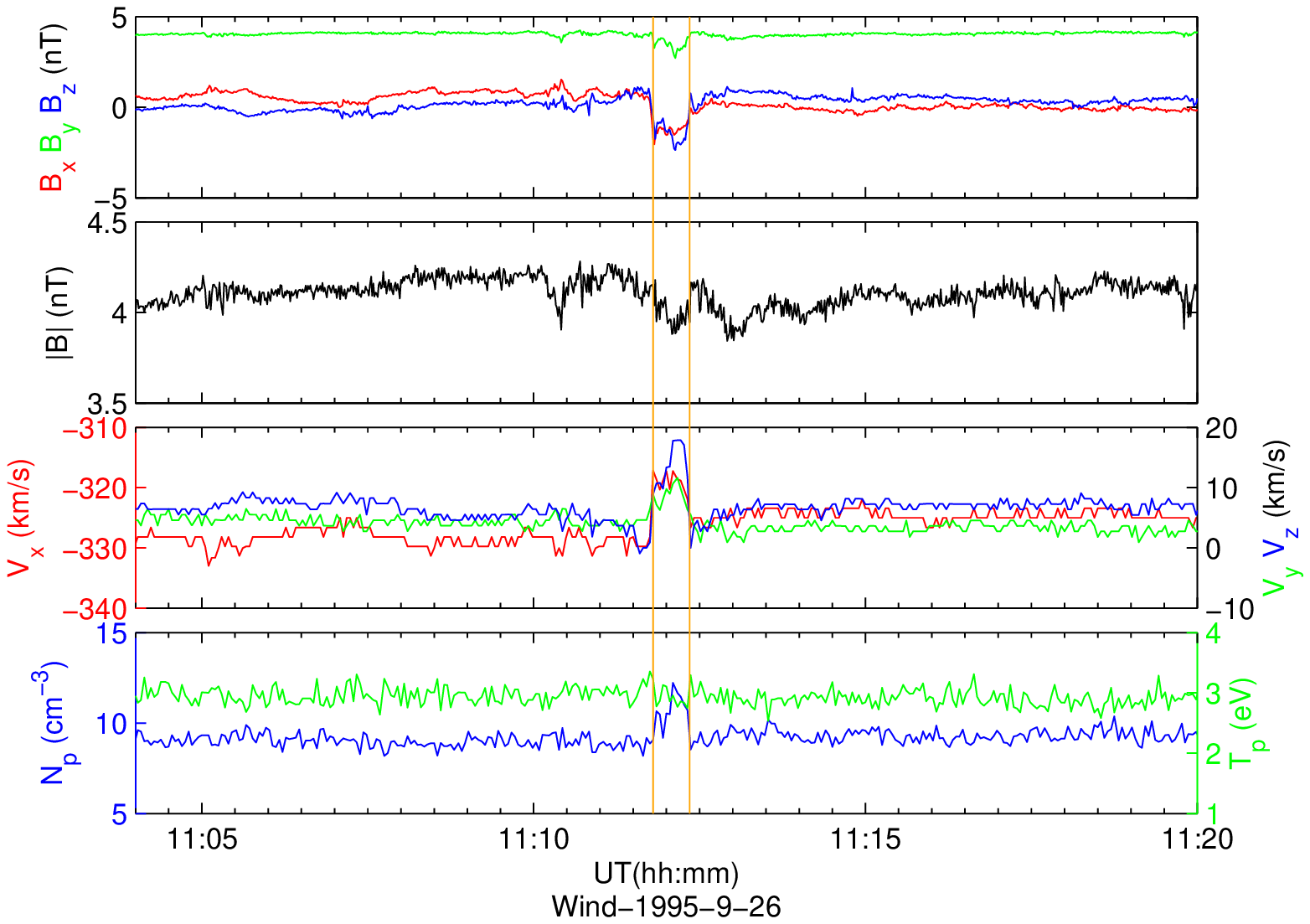}
\hspace{2in}
\caption{A double-current-sheet event that occurred in 1995-09-26. Shown from top to bottom are 
the $3$ components of the vector magnetic field in the Geocentric Solar Ecliptic (GSE)
coordinate system, the magnitude of magnetic field, the  $3$ components of the vector proton velocity 
in the Geocentric Solar Ecliptic (GSE) coordinate system, and the solar wind proton number density and proton 
temperature, respectively. The two brown vertical lines mark the location of the current sheet.}
\label{AnglePlot2}
\end{figure}

Figure~\ref{AnglePlot2} is a double-current-sheet event. Two current sheets can be identified around 
11:11:48 UT and 11:12:26 UT. In the scenario of flux-tube-crossing,
 Figure~\ref{AnglePlot2} can be understood as the spacecraft briefly crosses the magnetic wall between two 
flux tubes and then returns to the original flux tube.
The schematic of this event is shown in the second panel of Figure~\ref{Cartoon}.
 Note that the temperature decreases and the proton density increases between the two current sheets.
As in Figure~\ref{AnglePlot1},  although Figure~\ref{AnglePlot2} can be explained by the 
flux-tube-crossing scenario, it need not to be.
The angles between $\delta B$ and $\delta V$ across the two current sheets are $175$ and $167$ degrees. 
Unlike the first event, this double-current-sheet is not associated with a reconnection exhaust.
 {There was a slight but insignificant drop of the magnetic field magnitude, so it is unlikely to be 
a magnetic hole. The proton number density and proton temperature were also 
changed slightly at the two current sheets. These slight changes, together with the pulse-like changes 
of the $3$ magnetic field components, suggest that this structure could be 
pulsed Alfv\'{e}n wave \citep{Gosling.etal11,Gosling.etal12}. Note, if this was a pulsed Alfv\'{e}n wave,
then according to Figure 3(a) of \citet{Gosling.etal12} and the fact that it has a duration of $46$ seconds, 
 it would be a long duration pulsed Alfv\'{e}n wave.}
		
\begin{figure}[ht]
\includegraphics[width=0.9\textwidth]{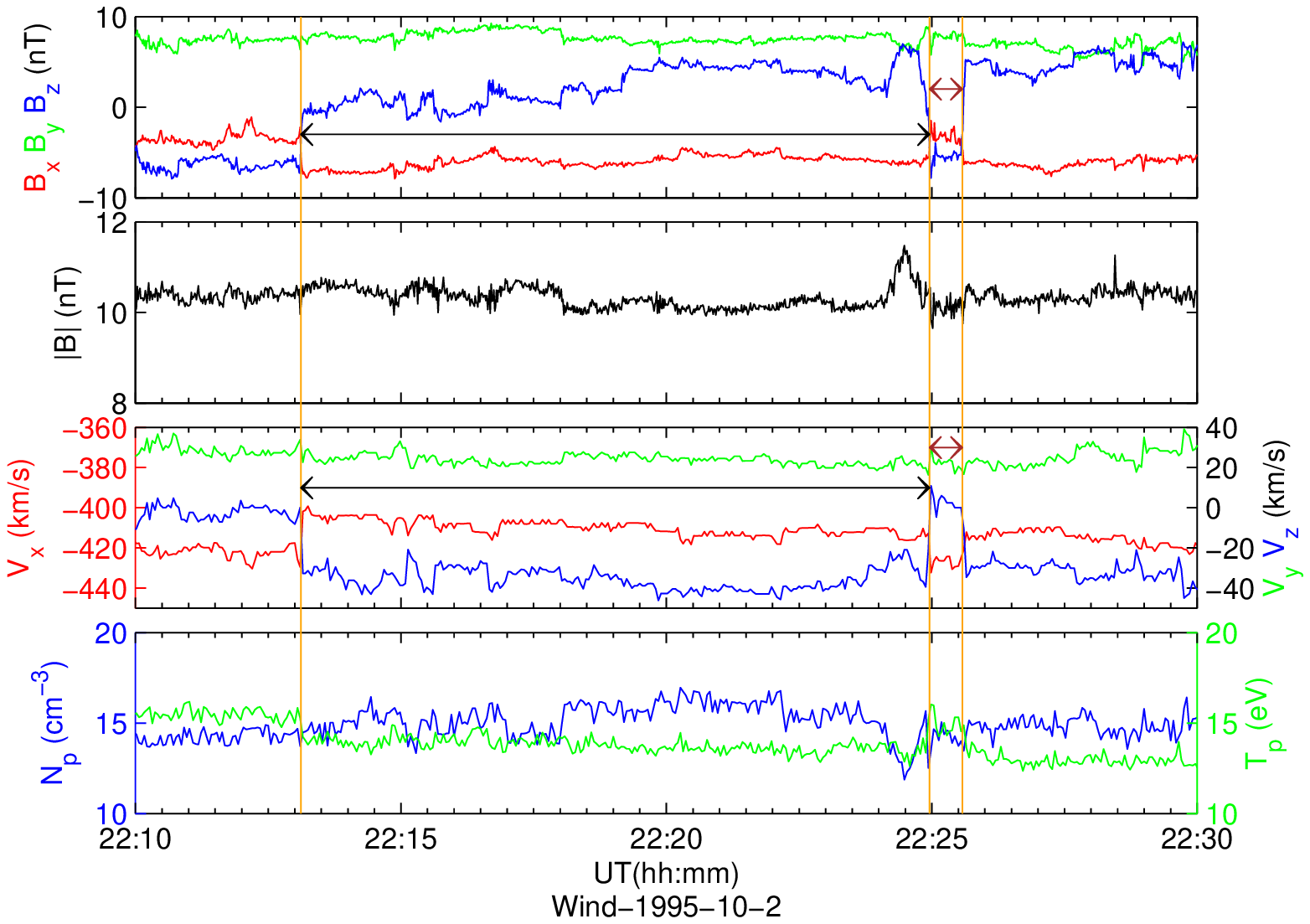}
\caption{A triple-current-sheet event occurred on 1995-10-02. 
Shown from top to bottom are the $3$ components of the vector magnetic field in the Geocentric Solar Ecliptic (GSE)
coordinate system, the magnitude of magnetic field, the  $3$ components of the vector proton velocity 
in the Geocentric Solar Ecliptic (GSE) coordinate system, and the solar wind proton number density and proton 
temperature, respectively. The three vertical lines mark the location of the current sheet.
Also see the online animation of the evolution of the unit magnetic field $\hat{B}$ in this event.}
\label{AnglePlot3}
\end{figure}

\begin{figure}[ht]
\includegraphics[width=0.9\textwidth]{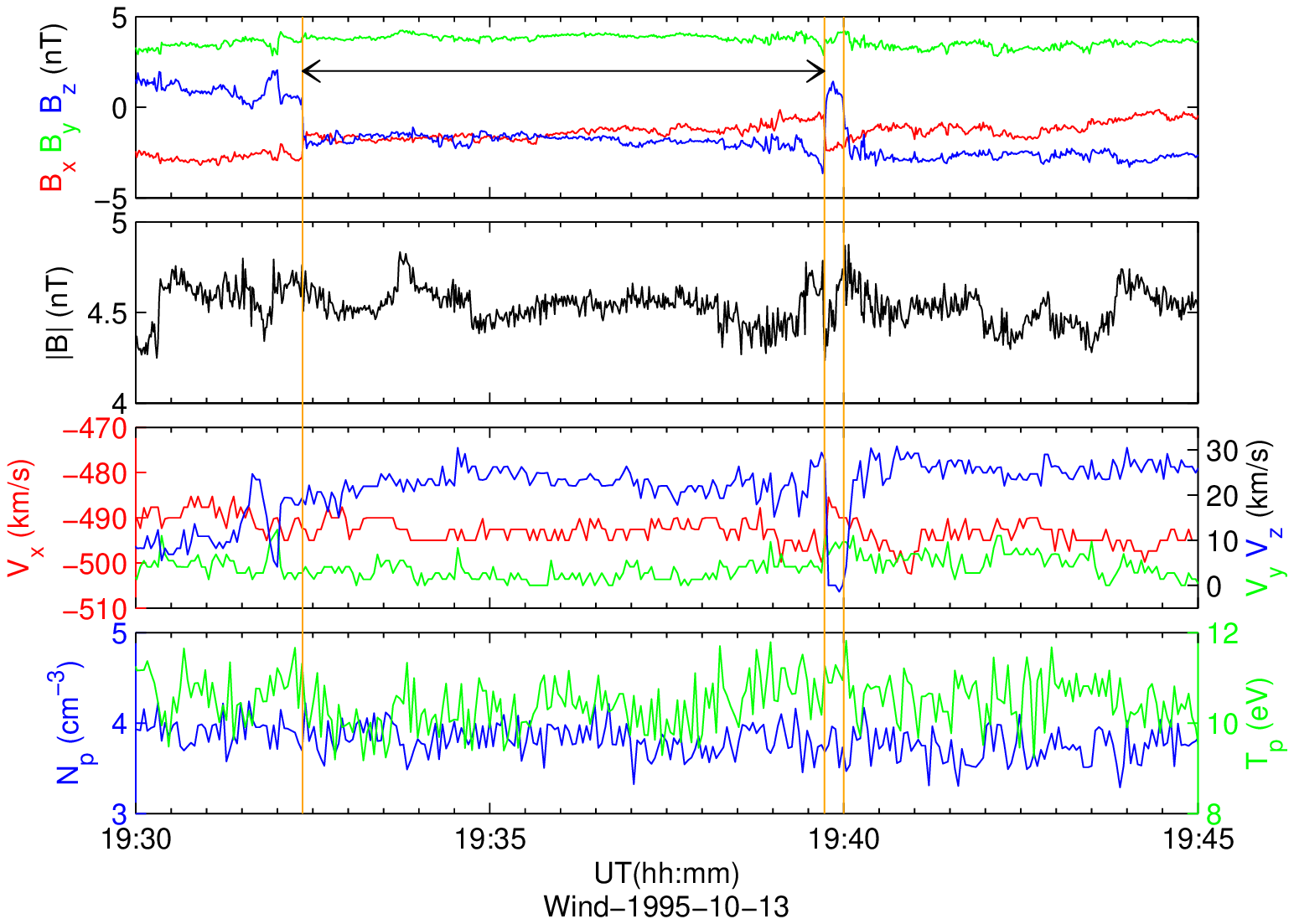}
\caption{Same as Figure~\ref{AnglePlot3}, but for the 1995-10-13 event. 
Shown from top to bottom are the $3$ components of the vector magnetic field in the Geocentric Solar Ecliptic (GSE)
coordinate system, the magnitude of magnetic field, the  $3$ components of the vector proton velocity 
in the Geocentric Solar Ecliptic (GSE) coordinate system, and the solar wind proton number density and proton 
temperature, respectively. The three vertical lines mark the location of the current sheet.
Also see the online animation of the evolution of the unit magnetic field $\hat{B}$ in this event.}
\label{AnglePlot4}
\end{figure}

Figure~\ref{AnglePlot3} and Figure~\ref{AnglePlot4} show two  triple-current-sheet  events 
that occurred on 1995-10-2 and 1995-10-13, respectively. Consider first the event shown in Figure~\ref{AnglePlot3}.
Throughout the event both $B_y$ and $V_y$ did not change much. 
$B_x$ underwent a sharp change at 22:13:05 UT, the first current sheet, 
after which it only changed slightly, until 22:25:00 UT when another 
sharp change occurred and $ B_x$ returned to values similar to those before 22:13:00 UT.
At 22:25:00 UT at the third current sheet another sharp change in Bx occurred.
After the crossing, $B_x$ at and after 22:25:32 UT returned to a value comparable to $B_x$ at 22:24:30 UT, 
just before crossing the 2nd current sheet. 
From the third panel, we can see that the $V_x$ changes at the same times as $B_x$. 

Similar behavior also occurred to $B_z$ and $V_z$. 
Before crossing the first current sheet at 22:13:05 UT, $B_z$ ($V_z$) was almost a constant.
After the crossing, $B_Z$ increased and $V_z$ decreased. $B_z$ also  
became slightly more turbulent.  At the 2nd crossing at 22:25:00 UT, 
$B_z$ and $V_z$ changed back to almost the same value as before 22:13:00 UT.
Then both $B_z$ and $V_z$ underwent another sudden change at the third current sheet crossing at 22:25:30 UT. 
After the third crossing, $B_z$ returned to a value similar to that before the 2nd crossing at 22:25:00 UT.

The magnitude of the magnetic field $|B| $ (the $2$-nd panel), the proton number density $N_p$
and the proton temperature $T_p$ (the $4$-th panel) did not vary much throughout the event. 
Before the crossing of the second current sheet, around 22:24:30 UT, 
$|B|$ increased and $N_{P}$ and $T_p$ decreased.  To better illustrate how the magnetic field direction 
evolves in this event, we have constructed an animation of the evolution of the unit magnetic field $\hat{B}$.

 {Two facts worth to note:
1) various plasma properties, including $N_p$, $T_p$ and the $3$ components of $\vec{B}$ and $\vec{V}$
in the short period between 22:25:00 UT and 22:25:30 UT are very similar to those prior to 
22:13:00 UT, suggesting that these are the same solar wind plasma. This can be 
clearly seen in the online animation.
2) similarly, the solar wind before and after the short period are likely the same and 
it is different from that in 1).} 

 {One may attempt to explain this triple-current-sheet event as the spacecraft crossing three 
uncorrelated individual current sheets that are generated by independent non-linear interactions of the 
solar wind MHD turbulence.  However, since independent current sheets have no correlations, 
the chance of the solar wind returning back to its original state after two independent current sheet 
crossings would be minute.} Alternatively, one may argue that the plasma between 22:13:05 UT and 22:25:00 UT 
represented a rather long-lived transient structure, and interpret the first two current sheets as the 
boundaries of this structure. In such a case, one has to explain why after the 
third current sheet crossing, both the magnetic field and the plasma return to values the 
same as inside the transient structure. 

 {Another triple-current-sheet event is the 1995-10-13 event, which is shown in Figure~\ref{AnglePlot4}.
Unlike the 1995-10-02 event, the $3$-components of $\vec{B}$ and $\vec{V}$, in particular, $B_x$ and 
$V_x$, suffered some additional changes at and around the three current sheets, making 
the 1995-10-13 event less convincing than the 1995-10-02 event. }

The first current sheet located at 19:32:28 UT.
Both $B_x$ and $B_z$ showed a sudden jump across the first current sheet;
 $V_x$ and $V_z$ did not show significant changes.
$B_y$ and $V_y$ also did not vary across the first current sheet.  The current sheet is 
therefore non-Alfv\'{e}nic. After crossing the first current sheet, 
$B_z$ was almost a constant for the next $\sim 7$
minutes until 19:39:40 UT, where the second current sheet was encountered. It increased across the 
second current sheet to a value similar to those prior to the crossing of the first current sheet.
Comparing to $B_z$, $B_x$ was nearly constant after crossing the 1st current sheet for  $\sim 3$ minutes and then 
gradually increased until 19:39:00 UT, after which it increases noticeably before the second current sheet. 
Across the second current sheet, it dropped to a value similar to those prior to the crossing of the first 
current sheet. The third current sheet occurred at 19:40:15 UT. Across the third current sheet, 
there was a significant change of $B_z$ and a small change of $B_x$. 
The two black horizontal dashed lines indicate that $B_x$ ($B_z$) before the first current 
sheet was similar to  $B_x$ ($B_z$) between the second and the third current sheets. 
The two magenta horizontal dashed lines indicate that  $B_x$ ($B_z$) between the first 
and the second current sheets was similar to  $B_x$ ($B_z$) after the third current sheet. 
Note that the change of $B_x$ at the third current sheet was smaller than that at the second current sheet.
After the third current sheet,  $B_x$ kept increasing, 
until  19:40:30 UT.  The value of $B_x$ after 19:40:30 UT is similar to those before 19:39:00 UT.
As in the 1995-10-2 event, we also constructed an animation of the evolution of the unit magnetic 
field $\hat{B}$ for this event.

For the 1995-10-02 event, the angles between $\delta B$ and $\delta V$ across the three current sheets 
 are $179^{\circ}$, $176^{\circ}$, and $174^{\circ}$, respectively.
For the 1995-10-13 event, the angles between $\delta B$ and $\delta V$ are $155^{\circ}$,  $124^{\circ}$ 
and $173^{\circ}$, respectively. 
While the three current sheets in the 1995-10-02 event are highly Alfv\'{e}nic, those in the
1005-10-13 event  are not.

\section{Discussion and Summary}
Current sheets are ubiquitous in the solar wind.  They can be generated in-situ through 
non-linear interactions of the solar wind MHD turbulence \citep{Greco.etal08, Greco.etal09}, 
or represent the boundaries of flux tubes that originated at the Sun
 \citep{Bruno.etal01, Borovsky08, Li.etal08}. 
Appearing in pairs,  they could also be the boundaries 
of reconnection exhausts (\citep{Gosling11}). 

An intriguing question one may ask is: for any particular current sheet, can we identify how it originated?

{  If current sheets that are generated in-situ and those that are convected out from the Sun have similar 
properties (such as the spread angles, the current sheet width, etc), then discriminating these two 
scenarios can be hard. } 
However, as shown in the rightmost cartoon in Figure~\ref{Cartoon}, the presence of 
triple-current-sheet event provides a strong support to the flux-tube scenario. 
This is because in the  flux-tube scenario the plasma and field changes 
across the three current sheets  are intimately correlated: 
as the spacecraft crosses the three current sheets, the plasma  before the first crossing  and
 that between the 2nd and the 3rd crossing are the same; the plasma between the first and the second crossings 
and that after the third are the same. This is in stark contrast to the scenario where 
the current sheets are generated in-situ. In the latter scenario, the 
plasma changes at the three current sheets in a triple-current-sheet event need not match.

Note that the identification of a triple-current-sheet event does not tell us how many single-current-sheet events 
are due to flux-tube crossing. As discussed earlier, 
since a reconnection exhaust can be of large scale \citep{Gosling07},
some  single-current-sheet events we identify can be the boundaries of reconnection exhausts.
\Citet{Gosling10} identified an occurrence rate of $40$-$80$ reconnection events 
per month in solar minimum. 
 {In our study, we only consider current sheets which are 
abrupt (width $< $10 seconds) and whose spread angles are larger than $45^{\circ}$, }
we find about $350$ ``single-current-sheet'' events per month. 
Assuming $2*60=120$ are boundaries of reconnection exhausts,
then the rest are presumably either generated in-situ or are the boundaries of flux tubes. 
Assuming $80\%$ ($50\%$) of the rest are generated in-situ, then one gets about  $45$ ($115$) 
single-current-sheet events that are flux-tube crossings per month.  

{ If current sheets are boundaries of flux tubes that have a solar origin, e.g. super granules,
 then one may expect to find some statistical correlations between in-situ observations of current sheets
and solar observation of super granules. Indeed, \citet{Bruno.etal01} have suggested that the sizes of
the flux tubes, when tracing back to the solar surface, may correlate with the size of photospheric magnetic 
networks. In the work of \citet{Miao.etal11}, using Ulysses observation,  the distribution of the
waiting time statistics of the current sheets were obtained. Assuming these flux tubes do not split or merge during 
their propagation to $1$ AU, then one may expect such waiting time statistics resembles the distribution of the 
magnetic network sizes. Examining the waiting time statistics of current sheet,  and in particular, 
its dependence on heliocentric distance, and its correlation with supergranule size will be reported in future 
work.}

To conclude, we have examined $2$-month's worth solar wind data from the {\it Wind} spacecraft and identified two
 triple-current-sheet events. The sequence of the observed magnetic field and plasma data in these two 
events are in agreement with the scenario where current sheets are flux tube boundaries, as depicted 
in Figure~\ref{Cartoon}. 
Unambiguous identification of flux tubes in the solar wind is important because these structures present an 
additional source of solar wind MHD turbulence intermittency. They can affect the power spectrum of the 
solar wind MHD turbulence \citep{Li.etal11, Li.etal12} as well as affecting the transport of energetic particles 
in the solar wind \citep{Qin.Li08}.

\acknowledgments
We thank R.P. Lepping and R.P. Lin and the CDAWeb for making available the data used in this paper and the 
referee for very valuable suggestions. This work is supported in part by NSF grants ATM-0847719, AGS0962658,
AGS1135432 and NASA grant NNH07ZDA001N-HGI and NNX07AL52A.

\end{document}